\documentstyle[aps,twocolumn]{revtex}
\begin{document}
\preprint{Usach USACH-FM-00-03}
\def\D{{\cal D}}
\def\12{{1\over 2}}
\def\st{\displaystyle}
\title{SCATTERING IN THREE DIMENSIONAL EXTREMAL BLACK HOLES}
\author{ J. Gamboa\thanks{E-mail: jgamboa@lauca.usach.cl} and F. M\'endez\thanks{E-mail:
fmendez@lauca.usach.cl}}
\address{Departamento de F\'{\i}sica, Universidad de Santiago de Chile,
Casilla 307, Santiago 2, Chile}
\maketitle
\begin{abstract}

\end{abstract}
\narrowtext
\maketitle

\section {Introduction}
Twenty five years ago Hawking \cite{hawking} discovered that  black holes (BH) could
disappear by quantum effects, suggestin that in a full quantum theory of
gravity the singularities could be smoothed out in the same sense as, for instance, the self-
energy of the electron is smoothed out by renormalization in quantum electrodynamics.

However before disappearance and in the case of very small BH, quantum effects
produced by these objects could be important and its one presence in the universe
should be enough  to detect new physical effects such as quantum corrections to
scattering of particles, new effects produced by strong fields and so on.

The scattering of particles around BH is, nevertheless, very interesting because for $D=
4$ the classical absorption cross section is proportional to the BH area \cite{kaplan}
(for results in higher dimensions see \cite{gibbons}) and one does not know if this result could survive at quantum
level\footnote{ For a descrition of this problem see \cite{unruh} and \cite{varios}.}.

One can solve this problem, considering a particle moving around a very small BH. If
the Schwarzschild radius is much bigger than the Compton length, then the BH can be
considered as a classical effective potential and the wave equation
\begin{equation}
\bigg[ \frac{1}{\sqrt{-g}} \partial_\mu (\sqrt{-g} g^{\mu\nu}
\partial_\nu)
- m^2\bigg] \psi = 0,
\label{kg}
\end{equation}
can be written as (assuming spherical or axial symmetry)
\begin{equation}
\frac{d^2R}{dr^2} + ( k^2  - V_{eff} (r) ) R = 0,
\label{radial}
\end{equation}
where $V_{eff}$ is the effective potential produced by the BH background. Module
technical refinements, this is the point of view followed in the literature \cite{birrel}.

In the last years, intense research in lower dimensional gravity has been performed and,
particularly, in three dimensions several classical and quantal results have been found.
From the quantum mechanics point of view  the equivalence between Chern-Simons
theories and $3D$ quantum gravity have been established \cite{witten} and classically a
three dimensional BH solution ($3D$ BH) has been found \cite{btz}.

Although three dimensional gravity is not related to the physical world, these results could be considered as a
theoretical laboratory where the Hawking's conjecture concerning  violation of quantum mechanics and other
expected properties could be proven \cite{hawking1}.

The purpose of this paper is to study the scattering of particles by a three dimensional BH
in the extremal limit. This is interesting by several reasons.  Firstly one could
expect that an extremal BH could be considered as a point particle and the scattering of
particles by this BH when $-1<M<0$ ($M$ the mass of the BH) could be seen as a similar problem to the Aharanov-
Bohm effect
with the BH playing the role of the solenoid \cite{ab}.

Secondly, if one can understand  scattering theory on an AdS spacetime one should
be able also, to understand the technical and conceptual differences between the
scattering by $3D$ BH and the Aharonov-Bohm effect.  This is a quite non-trivial
problem.

However, in order to formulate this problem one should understand firstly how to define scattering processes in an
anti-Sitter (AdS) space and, consequently, how to define asymptotic states.
In previous works have been considered in   this problem has been partially considered and an expression for the
absorption cross section has
been given avoiding the formal definition of   scattering processes \cite{birmin} (see also \cite{natsu})\footnote{For
the non-extremal case including subtle points  see \cite{myung}. }.
More recently,  following the last developments in the AdS-CFT  correspondence \cite{ah}, new approaches to the
scattering theory in AdS have been proposed\cite{scatt}.

The main results that we will describe below are the following:

a) The wave equation is exactly solved. Our solution shows  that outside  the horizon
the optical theorem is satisfied and  unitarity is not violated. However inside  the
horizon the hamiltonian is not self-adjoint and there is not unitarity.

b) The total cross section  vanishes.

c) We find a set of appropriates variables where the scattering problem is  well
defined;  in some sense this variables are a sort of a reciprocal space.

In the next section we will exactly solve  the Klein-Gordon equation for extremal 3D BH and  compute the absorption
cross section by means of
\begin{equation}
\sigma_{abs} = \frac{J(r \rightarrow r_+ )}{J (r \rightarrow \infty)} = \frac{J_{abs}}{J_{\infty}},
\label{abs}
\end{equation}
where $r_+$ is the horizon of the $3D$ BH. We find that $J_{abs} =0$ implies $\sigma =0$, {\it i.e.} the extremal
$3D$ black hole is a transparent object. In section III we reinterpret the above problem in a space where is possible to
define {\it in} and {\it out}  states and the calculation of $\sigma_{abs}$ is performed as in the standard scattering
theory.

\section {Solution of Klein-Gordon equation and flux of particles}

In order to prove the results mentioned above let us consider the $3D$ BH solution
found in \cite{btz}
\begin{equation}
ds^2 = - (N^2 - r^2N^\phi ) dt^2 + N^{-2} dr^2 + r^2 d \phi^2 + 2r^2N^\phi
dt d\phi,
\label{btz}
\end{equation}
 where the lapse function $N^2$ and $N^\phi$ are defined as
\begin{eqnarray}
N^2 &=& -M + \frac{r^2}{l^2} + \frac{J^2}{4r^2}, \\
N^\phi &=& -\frac{J}{2r^2}. \label{N}
\end{eqnarray}
 Here, $M$ and $J$ are the mass and angular momentum of the BH and $l^{-1} =
 \sqrt{-\Lambda}$.

The lapse function vanishes for
\begin{equation}
r_\pm = l {\biggl[ \frac{M}{2}\biggl( 1 \pm \sqrt{1 -
\frac{J^2}{M^2l^2}}\biggr)\biggr]}^{\frac{1}{2}}.
\label{hori}
\end{equation}

The horizon is identified with $r_+$ and it will exist only if $M$ and $J$
satisfy the relations
\begin{equation}
M>0, \,\,\,\,\,\,\,\,\,\,\,\,\vert J \vert \leq Ml.
\label{cond}
\end{equation}

The extreme case corresponds to put $\vert J \vert = Ml$ and in such a case both roots in
(\ref{hori}) coincide.

Using these results one could write the Klein-Gordon equation and  study the
scattering problem as it is usually  realized in quantum mechanics. The equation  (\ref{kg}), in the above metric
(\ref{btz}) can be solved if we consider the ansatz
\begin{equation}
\psi =  e^{i\omega t} e^{in\phi} f_{n\omega}(r),
\label{ans}
\end{equation}
 where $f_{n\omega} (r)$ is an unknown function. Once (\ref{ans}) is replaced in
 (\ref{kg}) one finds
 \begin{eqnarray}
 & &\biggl( \frac{(r^2 - r_+^2)^4}{r^2 l^4}\biggr) f''_{n \omega}(r) + \biggl( \frac{(r^2 - r_+^2)^3}{r^3 l^4}
 \biggr) (3r^2-r_+^2)f'_{n\omega}(r)+\nonumber \\
& & [(\omega l +n)((\omega l - n) r^2+2nr_+^2)-\frac{m^2}{l^2}(r^2-r_+^2)^2]f_{n\omega}(r) =0. \nonumber \\
 & &
 \label{kg2}
 \end{eqnarray}

In order to solve this equation, let us make the change of variable
\begin{equation}
\xi = \frac{r^2_+}{r^2 - r^2_+}.
\label{cv}
\end{equation}

Equation (\ref{kg2}) now becomes

\begin{equation}
{\cal L}f_{n\omega} (\xi) = 0, \label{kg3}
\end{equation}
 where the elliptic linear operator ${\cal L}$ is
\begin{equation}
{\cal L} = \frac{d^2}{d\xi^2} + k^2_0 + \frac{k_1}{\xi} -
\frac{k_2}{\xi^2}
\label{sch}
\end{equation}
with the constants defined as
\begin{eqnarray}
k^2_0 &=&\Omega_+^2, \label{en1}\\
k_1 &=& \Omega_+ \Omega_{-}, \label{en2} \\
k_2 &=& \frac{m^2}{4}.\label{en3}
\end{eqnarray}
where $\Omega_{\pm} = \frac{1}{\sqrt{2M}} (\omega \pm n)$ (we have put $l =1$).

Equation (\ref{kg3}) can be solved by standard methods \cite{courant}. The two linearly independent solutions are
\begin{eqnarray}
f_{n\omega}^{(1)} (\xi) &=&  e^{- i \Omega_+  \xi} \xi^{s_+}
F[ s_+ + i \frac{\Omega_-}{2} , 2s_+, 2 i \Omega_+ \xi],
\label{sol1} \\
f_{n\omega}^{(2)} (\xi) &=& e^{- i \Omega_+  \xi} \xi^{s_-}
F[ s_- + i \frac{\Omega_-}{2} , 2s_-, 2 i \Omega_+ \xi]. \nonumber \\
& & \label{sol2}
\end{eqnarray}
where $s_\pm = \frac{1}{2} (1 \pm \sqrt{1+m^2})$  and $F[a,c,z]$ is the confluent hypergeometric function (Kummer's solution).

In  equation (\ref{cv}) we can see that the region $r \rightarrow \infty$ corresponds  to $\xi \rightarrow 0$,  and
$r\rightarrow r_+$, to $\xi \rightarrow \infty$. Thus, the regular solutions are
\begin{eqnarray}
f_{n\omega}^{(1)} (\xi), & & \,\,\,\,\ r \rightarrow \infty\,\,\,  (\xi \rightarrow 0), \\
f_{n\omega}^{(2)} (\xi), & & \,\,\,\,\ r \rightarrow r_+  \,\,\,(\xi \rightarrow \infty).
\label{converg}
\end{eqnarray}

In order to compute the absorption cross section \cite{unruh,natsu,maldacena}, we need to know the flux of particles
in the horizon and infinity.  The conserved radial flux for (\ref{kg2}) in $\xi$ coordinates is
\begin{equation}
j_r(\xi)= f_{\omega n}^{*} (\xi) \frac{d}{d \xi} f_{\omega n} (\xi) - f_{\omega n} (\xi) \frac{d }{d \xi}f_{\omega
n}^{*} (\xi).
\label{flux}
\end{equation}

The solution of (\ref{kg2}) is
\begin{equation}
f_{n\omega}(\xi) = A f_{n\omega}^{(1)} (\xi) + B f_{n\omega}^{(2)} (\xi),
\label{gen}
\end{equation}
but near the horizon $f_{n\omega}^{(1)} (\xi)$ goes to infinity and, in order to have a regular solution, one can
choose  $A=0$ . Thus,  the  flux of particles in the horizon comes from the regular part of (\ref{gen}). The
calculation is straightforward if we  use the  Kummer's transformation  for the confluent hypergeometric function
$F(a,c;\xi)$,  {\it
i.e.}
\begin{equation}
F(a,c;\xi) = e^{\xi}F(c-a,c;-\xi).
\label{kumer}
\end{equation}
Using this fact, one finds - for regular solutions in the horizon - that  $j_{r} =0$.

On the other hand since  we are sending particles from infinity, $J_{\infty} \neq 0$ and consecuently  the absorption
cross section\footnote{Here the reader should note that this result is {\em independent} of the normalization
constants that appears in the solutions of (\ref{kg2}).} is
\begin{equation}
\sigma = \frac{j_{r_+}}{j_{\infty}}=0.
\label{sigma}
\end{equation}

Therefore, one finds that the total cross section  vanishes and this extremal BH is transparent for {\it any} energy of
particles. A similar  phenomenon occurs in  nature for the scattering of electrons by  inert atoms  where, for some
values of the energy, there is not scattering for $s$-waves, a phenomenon  known as Ramsauer-Townsend effect
\cite{davydov}.

Thus, our calculation shows that the quantum scattering of particles with $3D$ extremal
BH is always governed by a kind of Ramsauer-Townsend mechanism.


\section{Reciprocal Space}

Following the arguments given in section I, in an AdS space is not possible to define {\it in} and {\it out} states and,
as a consequence, the definition of the scattering theory is more subtle. In this section we will reinterpret the previous
results in order to find an alternative definition of the asymptotic states.

Indeed, equation (\ref{kg3}) can formally be seen as a Schr\"odinger equation in the potential
\begin{equation}
V(\xi) =  \frac{k_1}{\xi} - \frac{k_2}{\xi^2},
\label{pot}
\end{equation}
with $k^2_0$ playing the role of the energy and \lq \lq $\xi$ " a radial coordinate. Of course, the next question is:
what is the angle in this space?. As we are interested in the calculation of the total cross section, this angle can be
chosen equal to the $\phi$ angle that appears in (\ref{btz}), {\it i.e.} taking values on $0<\phi<2\pi$. We will call
${\cal H} = (\xi, \phi)$ the reciprocal space.

Following  general properties of elliptic differential equations
\cite{courant}, the eigenfunctions $f_{n\omega}$ should be continuous
everywhere, but as (\ref{pot}) is singular in $\xi = 0$, one must add an
additional condition on $f_{n\omega}$, {\it v.i.z.}
\begin{equation}
f_{n\omega} (0) = 0,
\label{reg}
\end{equation}
one could note that (\ref{reg}) is self-adjoint condition for the operator ${\cal L}$ \cite{govi}.

This last condition assure us continuity in $\xi = 0$ and the vanishing
of $V(\xi)$ when $\xi$ goes to infinity would permit to define asymptotic
states as in the usual scattering theory, {\it i.e.} the existence of a
general asymptotic solution of (\ref{kg3}) like $A(\phi) e^{ik_0 \xi}$,
where $A(\phi)$ and $e^{ik_0 \xi}$ are the scattering
amplitude and the asymptotic states, respectively. This last statement is
heavily dependent on $k_0$ being a positive real number. Looking at
(\ref{en1}), this mean of course that the inequality
\begin{equation}
{[ n + \omega ]}^2 > 0,
\label{ine}
\end{equation}
is verified.

One should also point out  that (\ref{reg}) implies a physical condition on
the system. If $\xi$ takes values on $\Re^{+}$ -- {\it i.e.} if
$\xi$ is interpreted as a
radial coordinate~-- ~(\ref{reg}) can be seen as a condition of
probability
conservation in the origin. By the contrary, if $\xi$ takes positive and
negative values ~-- ~{\it i.e.} just when the particles cross the horizon,
probability is not conserved and  unitarity is violated \cite{hawking1}.

Using (\ref{sol1}), the calculation of the scattering cross section is straightforward. Indeed, the asymptotic behavior
of $F$ for $\xi >>1$ contains a term like $1/\xi$ plus distorted waves corrections as in the Coulomb scattering.
Nevertheless,  the angular corrections only contain $e^{in\phi}$ and, as a consequence, the total scattering amplitude
simply  becomes
\begin{equation}
A(\phi) = \sum_{n= -\infty}^{\infty} e^{in\phi}, \label{am}
\end{equation}
which can be computed using the regularization prescription (see e.g.\cite{jackiw})
\begin{eqnarray}
A(\phi)  &= & 1 +\lim_{\epsilon \rightarrow 0}  \sum_{1} ^ {\infty} (e^{in(\phi + i \epsilon)} + e^{-in(\phi - i
\epsilon) } ), \nonumber \\
&=& 1 + 2 \pi \delta(\phi). \label{amp}
\end{eqnarray}

The optical theorem gives
\begin{equation}
\sigma \sim Im A(0 ) = 0. \label{opt}
\end{equation}

Therefore, in the $(\xi, \phi)$ reciprocal space we recuperate the Ramsauer-Townsend effect found  in section II.
Thus, one can assert that both descriptions are equivalents.

In conclusion, a) we have shown by two differents methods that the absorption cross section for spinless relativistic
particles in a $3D$ extremal BH background  vanishes  and  b) we have found an alternative description for the
scattering of particles in AdS spaces. The next step is to show the equivalence between our approach and the results
found in \cite{scatt}.

\pagebreak
\centerline{\bf Appendix}
\medskip
In this section we will show that the radial flux of the Klein-Gordon equation is  identically equals to
zero in the 3D black hole  background. The argument is as follows; noticing that the solution (\ref{sol2}) writen in
terms of $r$ also satisfy the property
\begin{equation}
f^{(2)*}(r) = f^{(2)}(r),
\label{ap1}
\end{equation}
for $r = r_+ +\epsilon$, then the flux
\begin{equation}
j_r = f^* (r N^2) \partial_r f - f  (r N^2) \partial_r f^*,
\label{flux}
\end{equation}
vanishes identically, independently of $\epsilon$. For incoming particles the incident flux $j_\infty$ is different from
zero by definition, otherwise there are no particles (this point is also discused in \cite{birmin}).

For the massless particle  one find the following solutions
\begin{eqnarray}
f_{n\omega}^{(1)} (\xi) &=&  e^{- i \Omega_+  \xi} \xi
F[ 1 + i \frac{\Omega_-}{2} , 2, 2 i \Omega_+ \xi],
\label{soll0} \\
f_{n\omega}^{(2)} (\xi) &=& e^{- i \Omega_+  \xi}
F[  i \frac{\Omega_-}{2} , 0, 2 i \Omega_+ \xi]. \nonumber \\
& & \label{soll}
\end{eqnarray}
but for the Kummer's relation $F(a,0;\xi] = e^\xi F(-a,0;\xi]$ and (\ref{ap1}) again is satisfied and the previous
results are obtained, {\it i.e.} $\sigma =0$.

\acknowledgments

 We would like to thank M. Ba\~nados  for  useful discussions. This work was partially supported by grants 1980788,
 3000005 from FONDECYT (Chile) and DICYT (USACH).

\end{document}